\documentclass[11pt,twoside]{article}

\usepackage{asp2004}
\usepackage{epsf}
\usepackage{lscape}
\usepackage{graphicx}

\begin{document}
\title{The $z\sim3$ QSO Luminosity Function with SWIRE}   
\author{B. Siana\altaffilmark{1}, M. Polletta$^2$, H.E. Smith$^2$, C.J. Lonsdale$^3$, E. Gonzalez-Solares$^4$, D. Farrah$^5$, T. S. R. Babbedge$^6$, M. Rowan-Robinson$^6$, J. Surace$^1$, D. Shupe$^1$, F. Fang$^1$}   
\affil{$^1$Spitzer Science Center, California Institute of Technology, Mail Code 220-6, 1200 East California Blvd., Pasadena, CA 91125 USA, $^2$UCSD, $^3$IPAC, $^4$Cambridge, $^5$Cornell, $^6$Imperial}    

\begin{abstract} 
We use a simple optical/infrared photometric selection of high redshift QSOs which identifies a Lyman Break in the optical and requires a red IRAC color to distinguish QSOs from common interlopers.  We find 100 $U$-dropout ($z\sim3$) QSO candidates with $r'<22$ within 11.2 deg$^2$ in the ELAIS-N1 \& ELAIS-N2 fields in the {\it Spitzer} Wide-area Infrared Extragalactic (SWIRE) Legacy Survey.  Spectroscopy of 10 candidates shows that they are all QSOs with $2.83<z<3.44$.  We use detailed simulations which incorporate variations in QSO SEDs, IGM transmission and imaging depth to derive a completeness of 85-90\% between $3.0<z<3.4$.  The resulting luminosity function is two magnitudes fainter than SDSS and, when combined with those data, gives a faint end slope $\beta=1.62\pm0.18$, consistent with measurements at $z<2$ and steeper than initial measurements at the same redshift.  
\end{abstract}

\section{Introduction}   
The QSO luminosity function is typically fit to a double power law of the form

\begin{equation}
\Phi(M,z) = \frac{0.92\Phi(M^*)}{10^{0.4(-\alpha+1)(M-M^*)}+10^{0.4(-\beta+1)(M-M^*)}}
\label{eqn:qlf_mag}
\end{equation}

\noindent (in magnitudes) where $\alpha$ and $\beta$ are the bright and faint end slopes, respectively.  At high redshift ($z>2.5$), initial indications are that the shape of the QSO luminosity function (QLF) may be changing.  Hunt et al (2004)(hereafter, H04), in a search of faint QSOs at $z\sim3$, find a very shallow faint end slope $\beta=1.24\pm0.07$.  At the bright end, Richards et al. (2006, in press) show that the slope of the bright end is getting shallower with redshift between $2.5<z<5$.  Given the importance of QSO luminosity functions in constrainting galaxy evolution models, growth of super-massive black holes (SMBHs), and QSO contribution to the ionization of HI and HeII in the IGM, these initial indications of a changing QLF warrant larger, deeper surveys to better constrain the high redshift QLF.  

It is difficult to obtain a complete QSO luminosity function at $z>2$ for two reasons.  (1) The colors of QSOs at $2.5<z<3.5$ coincide with those of main sequence stars, resulting in unreliable samples.  Therefore, targeted follow-up spectroscopy or imaging in redder bands (near-IR) is usually needed to distinguish interlopers from QSOs.  (2)  Surveys of faint QSOs at these redshifts require fairly deep imaging ($r'\sim24$, for accurate point source determination) over wide areas ($>10$ deg$^2$) to find large numbers.  The SWIRE Legacy Survey (Lonsdale et al. 2003), with IRAC \& MIPS (3.6-160$\mu$m) imaging over 49 deg$^2$ removes both of these obstacles.
 
\section{Selection}
The number of Ly$\alpha$ absorbers increases rapidly toward higher redshift.  The Lyman line and continuum absorption from these systems significantly depresses the continuum of high redshift objects blueward of Ly$\alpha$ (1216\AA).  Our optical color selection looks for this flux decrement (amongst point sources) as it redshifts into the $U$-band at $z\sim2.8$, causing a red $U-g'$ color in a source with an otherwise blue continuum.  The color criteria were chosed to select a QSO with a 2$\sigma$ deviation in UV spectral slope and are given in Eq. 1 and plotted in Figure 1.  Typically, high redshift QSO surveys avoid the color-space of main sequence stars, but our selection will include many A \& F stars as we expect to remove these with infrared colors.  

\begin{equation}
U-g'\geq 0.33 \hspace{7mm} g'-r'\leq 1.0 \hspace{7mm} U-g'\geq 3.9(g'-r')-2.0
\end{equation}

As shown by Lacy et al. (2004) and Stern et al. (2005), AGN candidates can be selected by their red IRAC colors.  Unfortunately, we do not expect to detect the majority of $r\sim22$ QSOs at these redshifts in the two longest wavelength IRAC bands (5.8 \& 8.0$\mu$m).  Therefore, we only use channels 1 \& 2 (hereafter, IRAC1 \& IRAC2) since they are two magnitudes more sensitive given the same exposure time.  In Figure 1 the IRAC colors of all point sources with $r' < 22$ in EN1 are plotted, showing that the IRAC1-IRAC2 color is sufficient in distinguishing stars from AGN.  Therefore, our infrared criteria are simply 

\begin{equation}
IRAC1-IRAC2 > -0.15 (AB) \hspace{10mm} f_{\nu}(4.5\mu m)>10\mu Jy.
\end{equation} 

\noindent We do not expect this cut to significantly affect our completeness, as the colors of our QSO template stay far above this cut between $2.8<z<3.5$, as do 18 of 19 SDSS QSOs in the SWIRE survey.

\begin{figure}[t]
\plottwo{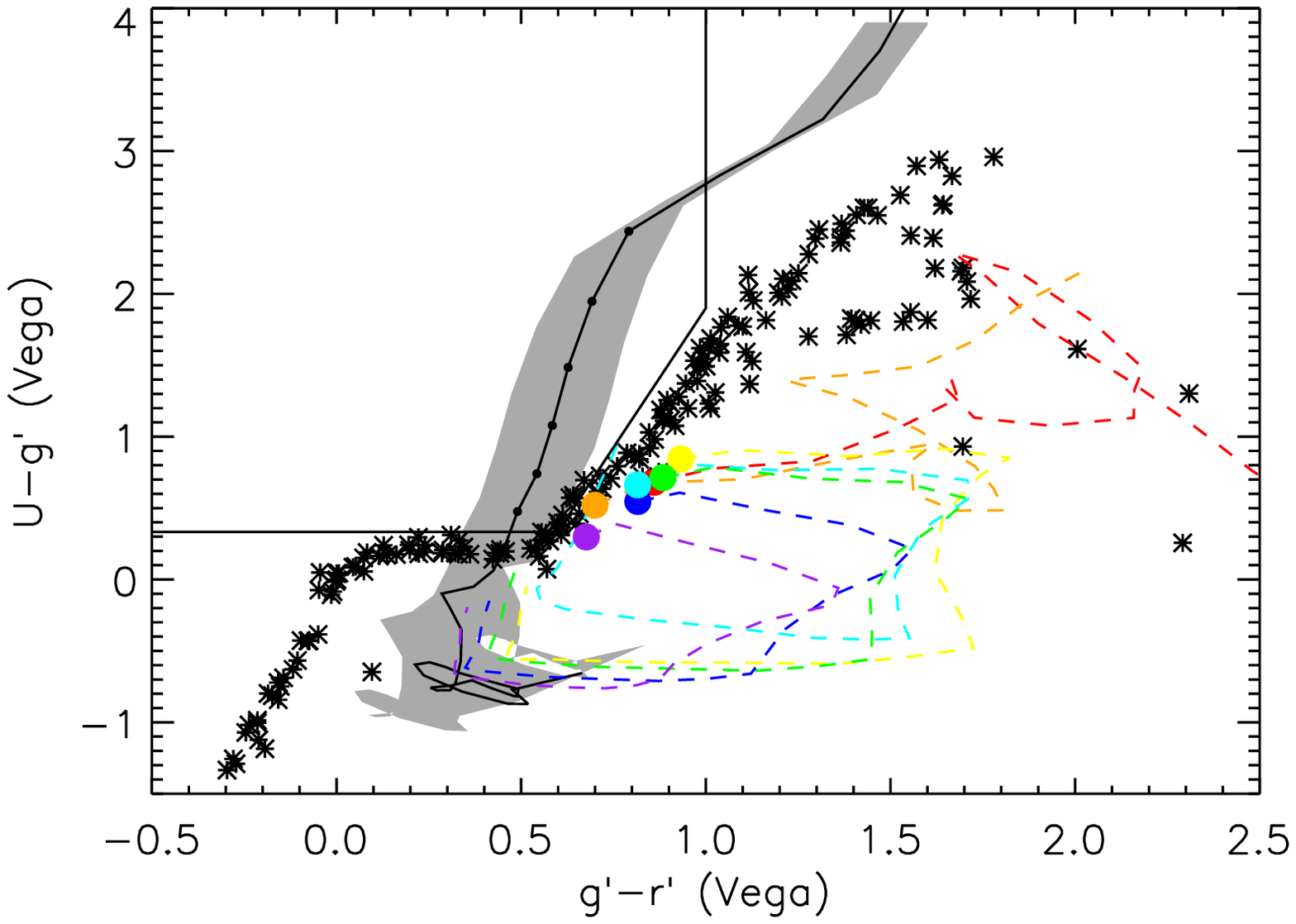}{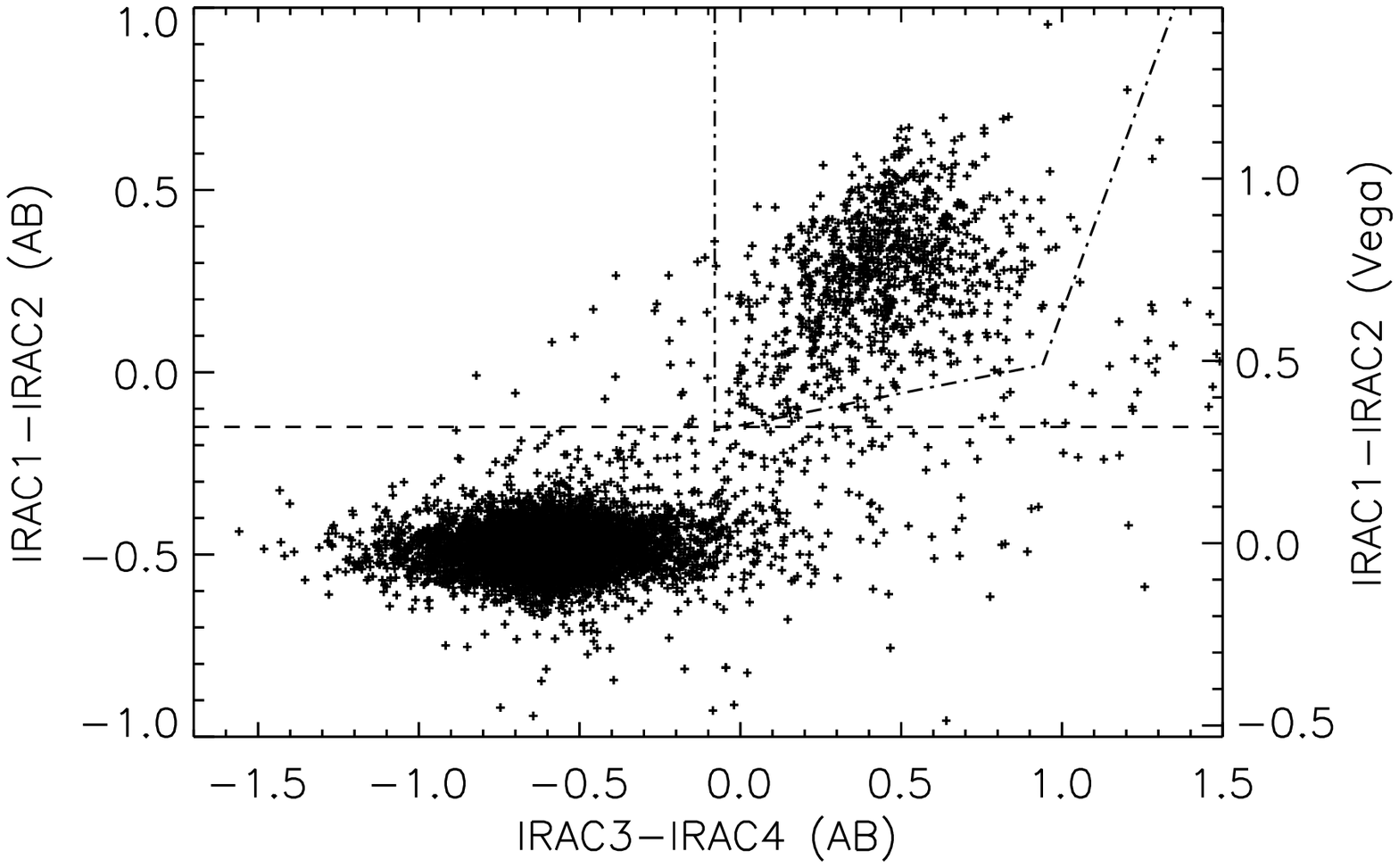}
\caption{Left panel: The optical color selection.  The black solid line is the QSO color track and the shaded area is the region occupied when shifting the QSO spectral slope by $\pm2\sigma$.  The QSOs are within our color selection between $2.9<z<3.4$.  The asterisks are main sequence stars (Gunn \& Stryker 1983) and the colored dashed lines are color tracks of various GRASIL galaxy templates (Silva et al. 1998) between $0<z<2$ .  Right Panel: The IRAC colors of point sources ($r'<22$) in EN1.  The AGN lie at the top right of the plot, while the stars occupy the lower left.  The AGN selection of Stern et al. (2005) (dash-dot) and uur infrared color cut (dashed) are plotted.}
\end{figure}
     
\section{Results}
We search the first two fields for which optical/infrared photometry was available (ELAIS-N1 \& ELAIS-N2).  These fields cover 11.2 deg$^2$ with both optical and infrared imaging, and yield 100 $U$-dropout candidates between $19<r'<22$.  Ten candidates were followed up with optical spectroscopy with Keck/LRIS and Palomar/COSMIC, confirming that all 10 are QSOs with $2.83<z<3.44$.  In addition, analysis of the complete optical-IR colors and four band IRAC colors (when available) confirm a reliability $>90$\%.  We perform detailed Monte-Carlo simulations which account for variations in the QSO spectral slope and emission line equivalent widths, the distribution of intervening Ly$\alpha$ absorbers, and imaging depth to derive the selection completeness,$C(r',z)$, as a function of redshift and magnitude.  We find that the completeness based on color selection alone is 85-90\% near the center of the redshift window. However, at the faintest magnitudes we are insensitive to some QSOs at the high redshift end ($3.3<z<3.5$) due to the limited depth our $U$-band imaging.  Therefore, the effective volume in the faintest bin is 72\% of the volume of the brightest bin.  We integrate the $C(r',z)$ function to derive an effective volume (and subsequent spece densities) for each magnitude bin. 

The binned QLF data are plotted in Figure 2 and match well with the faint end values from SDSS (Richards et al. 2006, in press).  We fit a double power-law to our data and the SDSS data and constrain the bright end slope to the value measured by SDSS ($\alpha=2.85$).  Our fitted parameters are $\beta=1.62\pm0.18$, $M^*$(1450\AA)$ = -25.6\pm0.3 (AB)$, and $\Phi^*=4.4\times10^{7}$ mag$^{-1}$ Mpc$^{-3}$ using a $\Lambda$CDM cosmology ($h=0.7$, $\Omega_m = 0.3$, $\Omega_{\Lambda}=0.7$).  

\section{Discussion}

Our fitted faint end slope is significantly steeper than that of H04.  A Kolmogorov-Smirnov (KS) test gives a probability P=0.004\% of measuring a more discrepant distribution from the same parent sample.  We therefore rule out, at the 4$\sigma$ level, the faint slope of H04.  We note, however, that at least part of the discrepancy may be explained by their use of only broad line ($FWHM>2000$km s$^{-1}$) AGN.  Although we expect most of our objects to have broad lines, we can not rule out a significant population of narrow line objects at the faint end of our sample.     The Pei (1995) fit has an almost identical faint end slope, but with a different normalization that predicts a factor of $\sim2$ more faint QSOs than our sample.  This may not be surprising since the shape of the Pei (1995) QLF was determined at low redshift and normalized with bright QSOs at high redshift.  Since the bright end slope gets shallower at high redshift, this causes the faint end of the Pei (1995) QLF to be too high.  We therefore see no indication of evolution of the faint end slope between $0<z<3.5$.  

\begin{figure}
\centering
\includegraphics[scale=0.6]{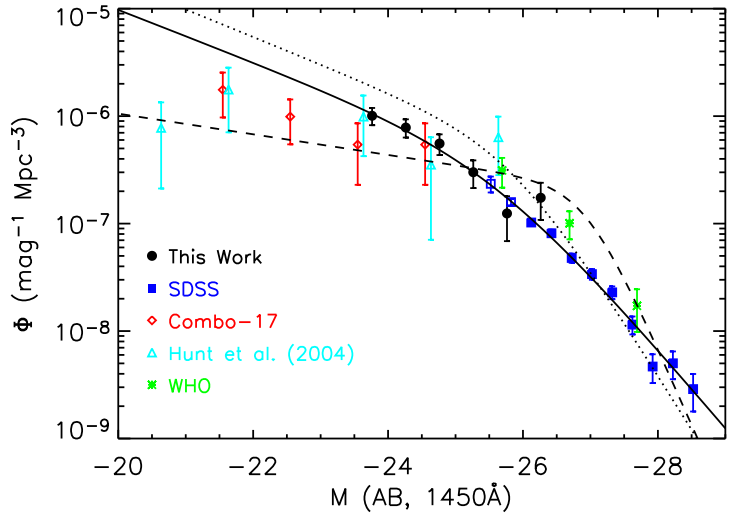}
\caption{The $z\sim3$ QSO luminosity function.  The binned data are plotted from this work (black circles), SDSS from Richards et al. (2006)(blue squares), Combo-17 (Wolf et al. 2003) (red diamonds), Hunt et al. (2004) (cyan triangles) and Warren et al. (1994)(WHO, green asterisks).  Filled symbols were used in the fitting of the QLF.  Also plotted are the fitted QLFs from this work (solid), Pei (1995)(dotted), and Hunt et al. (2004) (dashed).}
\end{figure}

We have also performed a $g'$-dropout ($z\sim4$) search, with spectroscopic follow-up of 10 candidates yielding 7 previously unknown $z\sim4$ QSOs.  However, three sources were galaxies at $z\sim0.4$, where the decrement in the $g'$ band was a Balmer or 4000\AA\ break.  Also, this sample suffered somewhwat from incompleteness as the redshifting of H$\alpha$ into IRAC1 at z=3.9 is expected to bring many of the QSOs below our IRAC color cut.  

For a more comprehensive analysis of the selection effects, comparison with other QLFs and implications for QSO contribution to HI photoionization at $z=3$, see Siana et al. (2006, submitted).



\end{document}